\documentclass[10pt,letterpaper]{article}
\usepackage[]{opex3}
\usepackage[]{graphicx}
\usepackage{dcolumn}
\usepackage{bm,multirow}
\usepackage{amsmath}
\usepackage{amssymb}

\begin{document}

\title{Design of an ultrahigh Quality factor silicon nitride 
photonic crystal nanocavity for coupling to diamond nanocrystals}

\author{Murray W. McCutcheon and Marko Lon\v{c}ar}
\address{School of Engineering and Applied Science, Harvard University,
Cambridge, MA, 02138}
\email{murray@seas.harvard.edu}

\date{\today}

\begin{abstract}
A photonic crystal nanocavity with a Quality ($Q$) factor of 
$2.3 \times 10^5$,
a mode volume of 0.55($\lambda/n$)$^3$, and an operating wavelength
of 637 nm is designed in a silicon nitride (SiN$_x$) ridge
waveguide with refractive index of 2.0.  The effect on the cavity
$Q$ factor and mode volume of single diamond nanocrystals of 
various sizes and locations embedded in the center and on top 
of the nanocavity is simulated, demonstrating that $Q > 2 \times 10^5$ is
achievable for realistic parameters.  An analysis of the figures of
merit for cavity quantum electrodynamics reveals that 
strong coupling between an embedded diamond nitrogen-vacancy center 
and the cavity mode is achievable for a range of cavity dimensions.
\end{abstract}



\section{Introduction}

Recently there has been much interest in solid-state approaches to the
study of quantum information, light-matter interactions and 
cavity quantum electrodynamics (QED)~\cite{Miller,Walther}.  
There are many potential advantages to implementing quantum protocols 
on a semiconductor chip.  The dipole coupling of matter to the field 
can be fixed 
because of the monolithic nature of the design.  Moreover, 
an integrated design can be naturally coupled to other on-chip devices,
both photonic and electronic, and is inherently scalable.
Strong coupling experiments in solid-state cavity 
QED have evolved from the 1D geometry of quantum wells in a Fabry Perot 
microcavity~\cite{Weisbuch92} to full three-dimensionally confining
micropillar and photonic crystal microcavities coupled to epitaxial 
quantum dots~\cite{Yoshie04, Reithmaier}.  More recently, there have been 
a number of significant advances~\cite{Hennessy07, Srinivasan07, Englund07} in 
both photonic crystal cavities and microdisks.  

Single nitrogen-vacancy (NV) defect centers in diamond
have recently emerged as promising candidates
for quantum optics and quantum information~\cite{Childress_Science06}.  
They act as
stable sources of single photons~\cite{Kurtsiefer}, and at room temperature,
they have electron spin coherence times of 350 $\mu$s~\cite{Gaebel06}
and nuclear spin coherence times of 0.5 ms.  
These spin states can be manipulated to form a quantum 
register~\cite{Dutt},
and coupled with reasonable strength to optical 
transitions such as the zero-phonon line (ZPL), allowing the read-out 
of the state~\cite{Santori,Hanson}.  NV centers occur naturally in bulk
diamond, but their spin states can dephase in the presence of
proximate nitrogen spins.  Recently, techniques have been developed to 
artificially implant NV centers in high-purity single crystal diamond, 
which limits the density of substitutional nitrogen atoms and the 
associated sources of decoherence~\cite{Gaebel06, Meijer}.  
Although much of this research relates to bulk diamond, 
NV centers embedded in 20 nm sized diamond 
nanocrystals (NCs) have also been shown to have spin 
coherence times on the order of microseconds~\cite{Jelezko04}.  
Embedding these diamond NCs in an optical
microcavity could allow realization of the coherent light-matter 
interactions crucial for certain quantum protocols.  With an 
appropriately designed cavity, for example, this coherence could be
controlled and entangled with photon states for transfer of the quantum
information~\cite{Greentree_PRA06, Cirac, vanEnk}.  
In fact, diamond NCs of about 75 nm in size have been strongly
coupled to the whispering gallery modes of silica microspheres~\cite{Park06}.

Our goal is to design a wavelength-scale microcavity for coupling to diamond
NCs on a planar
platform, which would facilitate integration with other optical devices.
A major challenge in realizing strongly-coupled diamond NV centers in this
system is the fact that the ZPL optical transition is in the visible (637 nm for 
the negatively charged NV$^-$ center).  A monolithic nanocavity must therefore
be designed in a visibly transparent material.  One option is diamond, which 
has a reasonably high refractive index $n=2.43$, and optical resonators 
have been designed for thin slabs of diamond~\cite{Tomljenovic,Kreuzer} with
$Q > 10^6$~\cite{Bayn}.  Experimentally this is very challenging, as the
growth of single crystal diamond slabs has yet to be realized, and the 
polycrystalline films which are readily available suffer from
large scattering losses which has limited the measured $Q$ factors to less 
than 1000~\cite{WangAPL07}.  In bulk single crystal diamond, moreover,
it is difficult to realize these structures due to the considerable
challenge of creating three-dimensionally confining defect cavities.

For material systems operating in the visible,
an alternative to diamond is a wide bandgap 
semiconductor, such as silicon nitride, hafnium oxide, gallium nitride, or 
gallium phosphide~\cite{Rivoire}.  Silicon nitride and hafnium oxide
are particularly 
promising candidates because of their compatibility with advanced silicon
nanofabrication processes.  Indeed, SiN$_x$ microdisks have been
fabricated with $Q$ = 3.6 $\times 10^6$ and mode volumes $V = 15
$($\lambda/n$)$^3$~\cite{Barclay06, Eichenfeld}.
The moderately low refractive index of $\sim$ 2.0 of SiN$_x$ has 
often been considered an impediment 
to ultra-high-$Q$ photonic crystal nanocavity designs, which thusfar
have only been demonstrated in high index semiconductors such as 
silicon~\cite{Noda05,Kuramochi06}.  To wit, the highest reported
photonic crystal cavity design in SiN$_x$ has yielded a $Q$ factor 
of 12,900 with a mode volume of 1.62 ($\lambda/n$)$^3$~\cite{Barth08}. 

In this paper, we demonstrate that silicon nitride photonic 
crystal nanocavities can have $Q$ factors of 230,000 with mode volumes of 
$\sim$ 0.55($\lambda/n$)$^3$. Considering the relative ease of 
fabrication and the natural integration of our design as part of an on-chip ridge 
waveguide, this remarkably high $Q/V$ ratio renders this device as a highly
promising platform on which to pursue visible solid-state cavity QED.
Although we focus on silicon nitride, these results are applicable for any 
low-loss material with n $\sim$ 2.0.

We first discuss our systematic approach to engineer the nanocavity,
which is based on a series of publications by Lalanne et 
al.~\cite{Lalanne_JQE, Lalanne_04, Sauvan}.
We then consider the effect on the mode $Q$ factor of a diamond NC
embedded in the center of the cavity or
positioned on top of the cavity surface, and demonstrate that a $Q$
factor greater than 95\% of the bare cavity $Q$ is obtainable 
for a realistic NC size of 20 nm.  For the case of the NC on top
of the cavity, we explore the effect
of spatial location of the NC with respect to the mode center, and 
evaluate the effect on the mode $Q$. We then
evaluate the cavity QED figures of merit and show that when the NC is
embedded in the center of the cavity, the system
is capable of realizing the strong coupling limit.

\section{Cavity design}

\begin{figure}[tb]
\centering
\vspace{10pt}
\includegraphics[width=10cm]{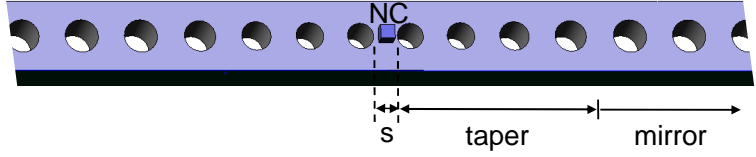}
\caption{Sketch of the 1D photonic crystal waveguide nanocavity, showing
a 60 nm diamond nanocrystal located on top of a cavity of length $s$.  
The taper and photonic crystal mirror sections are symmetric about the
cavity center.
}
\label{fig:3dview}
\end{figure}

Because of the relatively low index of refraction ($n$ = 2.0) of 
SiN$_x$, the two-dimensional photonic bandgap of SiN$_x$ planar photonic
crystal slabs is small, particularly when measured against
a high index semiconductor like silicon.  This makes the design of 
nanocavities in 2D photonic lattices challenging~\cite{Barth07, Barth08}.  
To circumvent this difficulty, we consider a nanocavity for which the
photonic lattice provides only one dimensional (1D) confinement, and 
total internal reflection provides the confinement mechanism in the
other two dimensions (2D).  In these ``1D + 2'' structures, the effective
1D bandgap, or stopband, is considerably larger than the corresponding
2D gap in a planar structure.  For example,
for the PhC considered in this paper, the stop-band of the
``1D + 2'' structure spans 13\% of the center frequency, whereas the
``2D + 1'' hexagonal planar photonic crystal with the same pitch
and hole radius yields a 2D bandgap spanning just 7\% of the 
center frequency.  The wider 1D stopband
allows a greater isolation of our defect mode from the band edges
and therefore higher $Q$ factors in engineered defect cavities. Recently,
a similar 1D design approach yielded a $Q$ factor of $2\times 10^8$ in a
high-index material ($n=3.46$)~\cite{Notomi_1D}.  A
sketch of the 1D nanocavity is shown in Figure~\ref{fig:3dview}, which
also shows a 60 nm diamond nanocrystal positioned on top of the cavity.

The design process consists of engineering three elements:  
(a) the photonic crystal mirror, (b) the taper, and (c)
the cavity length. We begin by considering a free-standing SiN$_x$ ridge
waveguide of thickness 200 nm and width 300 nm which supports
a single TE mode.  These parameters are chosen as representative of
our experimental wafer, and have not been
optimized in any fashion.  The photonic crystal parameters, namely
the hole spacing $a$ and radius $r$, are chosen to center the resulting
stop-band around the wavelength of interest ($\sim$ 637 nm).  Setting $a=250$
nm and $r/a = 0.28$ gives a stopband over the wavelength band 593-679 nm, 
with a mid-gap wavelength of 636 nm, as required.  We can characterize 
the reflectivity of the mirror using the 3D finite-difference time-domain
(FDTD) method by launching a 
waveguide mode pulse and monitoring the reflection spectrum.  
The reflected light is spatially integrated over
a plane perpendicular to the waveguide placed 300 nm in front of
the mirror, and the loss spectrum, defined as L = 1 - R, is calculated.  
This spectrum quantifies the amount of light which is not reflected by the 
mirror and is lost due to transmission and scattering losses.

As shown by Sauvan et al.~\cite{Sauvan}, the scattering losses arise 
in part from the mode mismatch between the waveguide and Bloch mirror 
modes.  By tapering the mirror, the discontinuity at the interface
between the cavity and mirror can be smoothed, allowing for an adiabatic 
transition between the modes in the two regions.  The mismatch can be 
quantified by comparing the effective index of the Bloch mode,
$\lambda/2a = n_{\rm Bl} = 1.274$, to the effective index of the
waveguide mode, $n_{\rm wg} = 1.480$ (determined using our 
3D-FDTD mode solver~\cite{Lumerical}).  
To gradually taper into the waveguide mode, the final
segment of the mirror should support a Bloch mode such that
$n_{\rm Bl} = n_{\rm eff} = 1.480$, which determines
$a_1 = 214$ nm.  For this to work, the $r/a$ ratio should be
maintained close to that of the uniform mirror, which implies $r_1 \sim 60$ nm.
Thus, we wish to smoothly taper the photonic crystal parameters ($a, r$) in the 
mirror from ($250, 70$) down to the values ($214, 60$) next to the cavity.  
In fact, we found an improvement by tapering
the radius to a smaller value than this rule would suggest, from 70 nm 
to 55 nm.

\begin{figure}[htb]
\centering
\includegraphics[width=10cm]{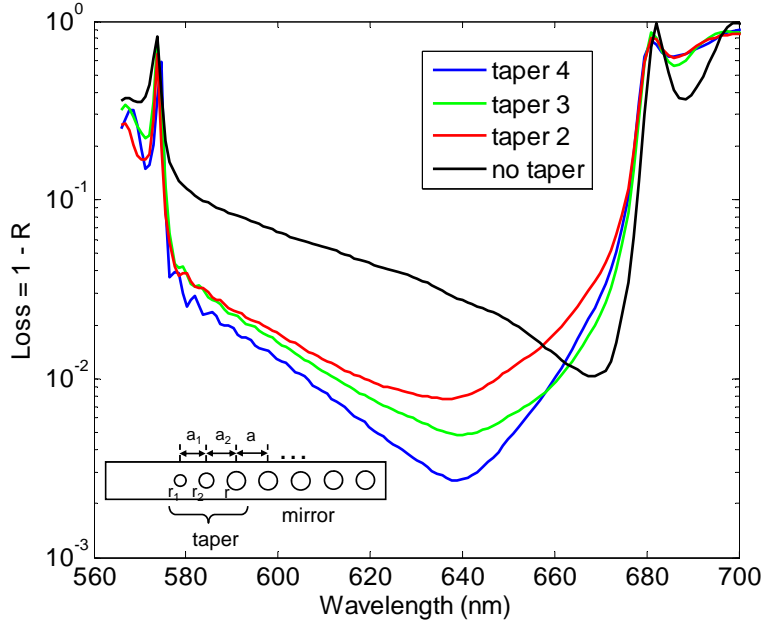}
\caption{Loss parameter for 4 different designs of photonic crystal
mirrors.  The $Q$ factors for the best nanocavities designed for
each taper are (13,000; 22,900; 230,000) for the (2; 3; 4)-section
tapers, respectively.  The inset defines the hole pitch ($a$) and 
radii ($r$).  The 4-hole taper parameters (in nm) are 
($r_1, r_2, r_3, r_4$) = (55, 58, 62, 66) and ($a_1, a_2, a_3,
a_4$) = (214, 226, 238, 250).
}
\label{fig:refloss}
\end{figure}

We gauge the efficacy of 2, 3, and 4 hole linear tapers according to the 
magnitude of $L$, as plotted in Figure~\ref{fig:refloss}.  The photonic 
crystal mirror is 17 periods in length, which was found to saturate the
in-waveguide $Q$ factor of the nanocavities based on these designs, as
discussed below. Each
increase in the taper length has the effect of further reducing the 
mismatch between mirror and waveguide.  The 4-hole taper 
yields a significantly improved reflectivity and hence lower loss, 
giving a minimum mirror loss $L < 0.003$.  

The nanocavity is defined by the gap $s$ between two taper/mirror
sections, as shown in Fig.~\ref{fig:3dview}.
In the 3D-FDTD simulation, the cavity is 
excited with several randomly phased and positioned dipole sources,
and the $Q$ factor is
determined from the exponential decay of the electric 
field ringdown in the cavity.  We also verified the $Q$ factors
by monitoring the power absorbed by the simulation 
boundaries, as discussed below, and these agree well with the ringdown 
values.  To accurately model in a 3D-FDTD simulation the exact taper 
and cavity length dimensions to the scale of nanometers 
while maintaining tractable simulation times, we use a graded mesh 
approach.  The mesh grid size is 5 nm in the central 0.4 $\times$
0.4 $\times$ 0.4 $\mu$m simulation volume.  Outside this 
region, the mesh grid spacing is 10 nm for the dielectric waveguide
 (corresponding to 25 
points per photonic lattice spacing), and is nonlinearly graded with
distance in free-space away from the dielectric material.  We verified
the convergence of the graded mesh simulation with a 5 nm grid uniformly 
meshed simulation.  

\begin{figure}[htb]
\centering
\includegraphics[width=10cm]{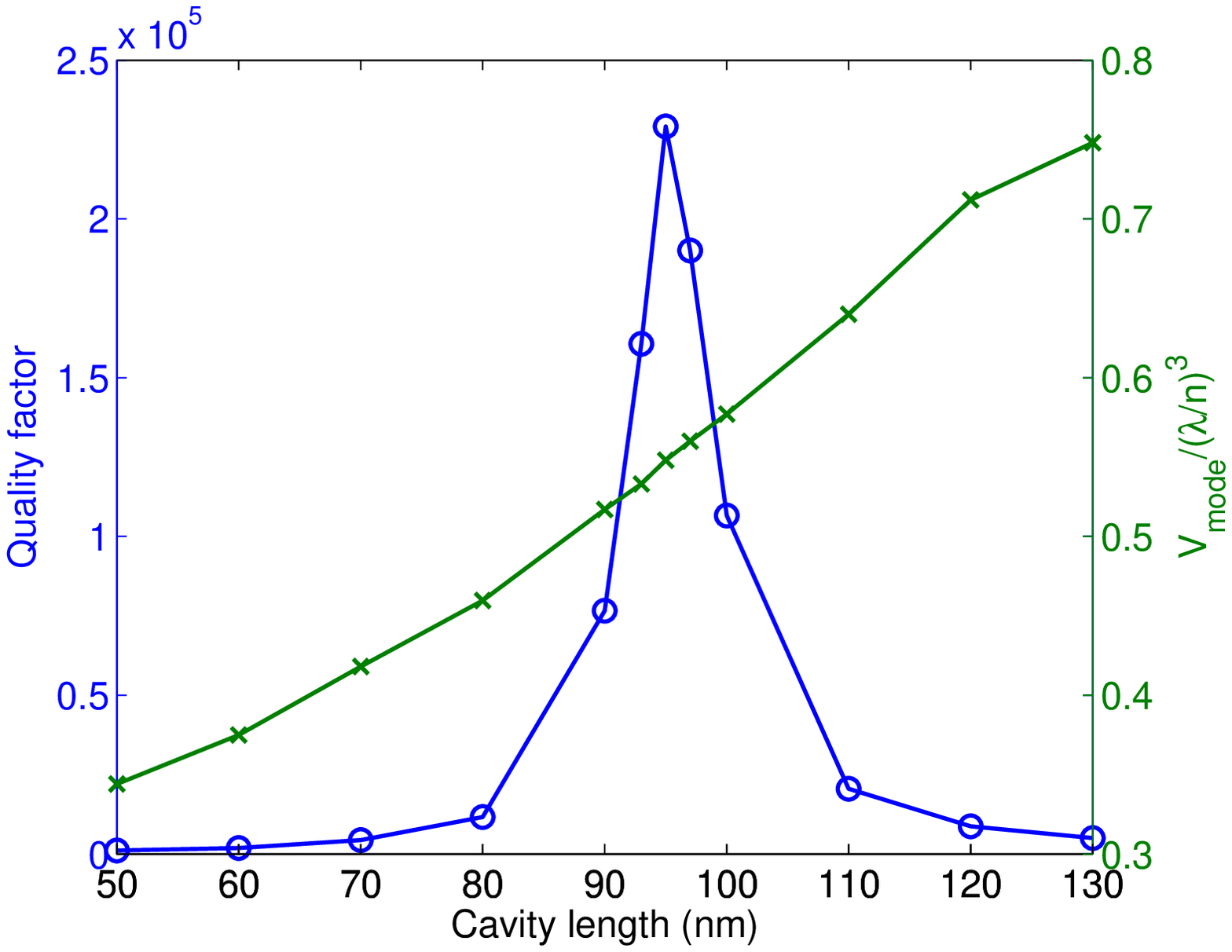}
\includegraphics[width=10cm]{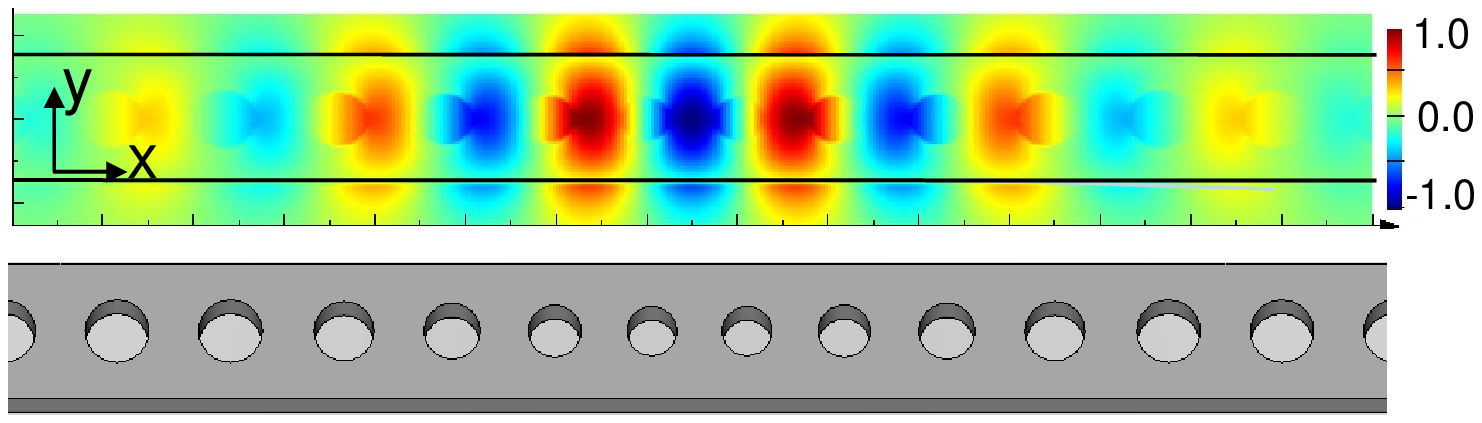}
\caption{Quality factor and effective mode volume as a function of
cavity length for a free-standing PhC ridge waveguide in SiN$_x$ with
a 4-hole taper and 17 period mirror.
The cavity mode electric field $E_y$ distribution is shown. }
\label{fig:QV}
\end{figure}

We consider
cavities formed by the 4-segment tapered mirrors, which gave the
highest reflectivity in the above analysis.  The dependence
of the $Q$ factor on the nanocavity length $s$ is illustrated
in Figure~\ref{fig:QV}.   The maximum $Q$ obtained
is 230,000 for $s = 95$ nm.
This is nearly 20 times better than the highest value
reported in the literature to date~\cite{Barth08}.  In these simulations,
we have not considered material absorption, which was shown by Barclay
et al.~\cite{Barclay06,Borselli05} to be a limiting factor in SiN$_x$
only for $Q$ 
factors in the range of a few million, or scattering due to fabrication
imperfections.  Further optimization may be possible by optimizing
the waveguide width and thickness, as well as using different (e.g.
longer) taper forms.

The mode volume is plotted as the green trace in Figure~\ref{fig:QV}.  
The cavity with the maximum $Q$ has a mode volume $\sim 0.55 (\lambda/n)^3$.
Despite the relatively low refractive index of SiN$_x$, this is half the 
effective mode volume of the recent ultra-high-$Q$ photonic crystal 
cavities designed in high index semiconductors
~\cite{Noda05, Kuramochi06, Notomi_1D},
demonstrating the highly attractive ultra-small size of these 1D 
nanocavities.

The mechanism which yields the
high $Q$ factor in the 4-hole tapered nanocavities is more subtle
than a simple reduction of mirror loss compared to the shorter tapers.  
As shown in Figure~\ref{fig:refloss}, 
the loss of the 2-hole taper is improved by a factor of $\sim$ 3
in the 4-hole taper, but the $Q$ factor of the optimal cavity based
on each design is increased by a factor of almost 20.  In a
Fabry-Perot cavity, one would expect the
$Q$ factor to be proportional to the inverse loss.  The difference
here is that the cavity mode is derived from the 
propagating Bloch modes of the photonic crystal mirror, and the
local structural perturbation which forms the cavity 
introduces a mode gap~\cite{Notomi_1D,Noda05} capable of supporting 
very high $Q$ modes.  Recently, we used a similar approach to design
cavities based on semiconductor nanowires with $Q \sim 10^6$~\cite{Zhang}.
The tapered transition
from the mirror to the nanocavity not only reduces mirror loss in the
simple Fabry-Perot picture, but reduces radiative loss arising from
delocalization of the mode profile in $k$-space~\cite{Srinivasan_OE}.
There may also be a role played by radiation 
modes in recycling the mirror losses and increasing the $Q$ factor, as 
elaborated by Lalanne et al. in Ref.~\cite{Lalanne_04}.

For a given structure,
the $Q$ factor can be separated into in-waveguide, $Q_{\rm wg}$, and 
perpendicular, $Q_{\perp}$ components using the relation
$1/Q_T = 1/Q_{\rm wg} + 1/Q_{\perp}$ (in analogy to the
$Q_{\parallel}$, $Q_{\perp}$ division common in 
the analysis of 2D PhC microcavities~\cite{Painter99}).  $Q_{\rm wg}$ is
calculated by measuring the power absorbed at the boundary of the
simulation within one half-wavelength of the surface of the waveguide, 
$P_{\rm wg}$, and then using the definition of the Quality factor,
$Q_{\rm wg} = \omega W/P_{\rm wg}$, where $W$ is the electromagnetic energy
in the resonant mode. $Q_{\perp}$ is then determined from $Q_{\rm wg}$ and
the total $Q_T$.  For a 17 period mirror, $Q_T=Q_\perp$, indicating that the 
waveguide losses have saturated and the total losses are completely
determined by out-of-waveguide loss.  With a shorter photonic crystal
mirror, the total $Q$ drops as it is limited by in-waveguide losses.
With 11 periods in the mirror, $Q_T$ is reduced to about half the 
maximum value.  This provides a guide for the design of an 
integrated in-waveguide emitter, in which light trapped or emitted into 
the nanocavity would couple predominantly into the waveguide mode rather
than scattering off the ridge.  This was the approach
of Zain et al.~\cite{Zain}, who designed a $\sim 6$ period PhC mirror in
a silicon-on-insulator ridge waveguide and
experimentally demonstrated a $Q$ factor of 18,500 with a transmission
of 85\%.

\section{Incorporation of a diamond nanocrystal}

Now that we have optimized our photonic crystal nanocavity design, 
we consider exploiting this nanocavity to enhance the zero-phonon line
(ZPL) emission from a NV center in a diamond nanocrystal.  Coupling 
the emission
to the nanocavity mode will lead to a Purcell enhancement of the 
spontaneous emission rate, and as we show below, the potential to
realize quantum dynamics in the strong coupling regime.

The ideal placement of the diamond NC is in the middle of the nanocavity, where
the NV center can interact with the maximal electric field of the cavity 
mode.  It is important to quantify the effect of the NC on the cavity $Q$ 
factor.  As in the bare (unperturbed) cavity 3D-FDTD simulations, 
a 5 nm mesh is used in the central volume of the 
simulation in order to accurately model the effects of the small volume  
of diamond.  We consider 20 nm and 40 nm sized 
cubic diamond NCs.
Interestingly, the results reveal that the $Q$ factor is
not uniformly affected for all cavities.  For cavities longer than
95 nm incorporating a 20 nm NC,
the $Q$ decreases by 10-12\%, but for cavities shorter than this,
the $Q$ factor is actually {\em increased}.  In the best
cavity ($s$ = 95 nm), the bare $Q$ factor of 230,000 increases to 240,000 
with the 20 nm NC.  These results perhaps reflect a trade-off
between the beneficial impact of the increased refractive index in the 
cavity center, which would be expected to raise the $Q$ factor, and the 
deleterious effect of changing the impedance matching condition used to 
design the mirror tapers.  As expected with the higher index cavity core, 
there is a reduction of about 10\% in the mode volume for all cavity 
lengths. We conducted a similar
investigation for a 40 nm cubic diamond NC located at the cavity center, 
and found that the $Q$ vs. $s$ curve shifts to shorter 
cavities compared to Fig.~\ref{fig:QV}, yielding a peak $Q = 260,000$
at $s=90$ nm, with a mode volume of 0.50 ($\lambda/n$)$^3$.
The main conclusion that can be drawn from this analysis
is that the cavity $Q$ factors
are not changed by more than about 10\% from the bare cavity values.  The
highest $Q$ factor can still be achieved, and even slightly improved,
and the mode volumes are slightly reduced.

Experimentally, an embedded NC might be realized by depositing 
a 100 nm layer of SiN$_x$ on a sacrificial layer of SiO$_2$, 
placing a single diamond NC in a known position, and then 
depositing another 100 nm layer of SiN$_x$ 
to cap the structure and embed the NC in
the middle.  The NC position could be registered with respect to 
external alignment markers~\cite{Hennessy04}, or the position might be revealed
after the SiN$_x$ regrowth by a bump on the top surface~\cite{Badolato05}.
The PhC nanocavity would then be patterned around the NC, followed by removal
of the SiO$_2$.  We simulated
the scenario of a small 20 nm bump of SiN$_x$ on top of the cavity with
an embedded 20 nm NC, and found a reduction of only about 2\% in the $Q$ 
factor of the best cavity.

An alternate approach to embedding a diamond NC would be to position the
NC on top of the cavity.  In~\cite{Koenderink}, 
Koenderink et al. consider
the emission enhancement of a dipole right at the semiconductor/air
interface of a uniform slab 2D photonic crystal, and show that the
rate can be enhanced by 5 - 10 times near the band edges.  Here we
model a similar scenario, except the emitter is situated on the
top surface of our SiN$_x$ nanocavity.  The high $Q/V$ ratio of our
nanocavity presents the possibility to yield much higher emission
enhancements, as elaborated below, with a maximum Purcell factor 
of 1,800 for the NC on top, and 27,000 for the cavity with an embedded 
NC.

We consider diamond NC cubes with 3 different edge lengths:  20 nm, 
40 nm, and 60 nm, and consider the effect of such a NC placed
exactly above the central anti-node of the nanocavity mode.  We
explore the effect on the mode $Q$ factor of this 
small dielectric perturbation (n = 2.43), and analyze the sensitivity
of the $Q$ to the precise positioning of the NC.  The cavity design
for these simulations had a maximum $Q$ of 115,000 ($s = 100$ nm),
which was not the optimal design ($s = 95$ nm).

\begin{figure}[htbp]
\begin{center}
$\begin{array}{c@{\hspace{-.5cm}}c}
\includegraphics[width=5cm]{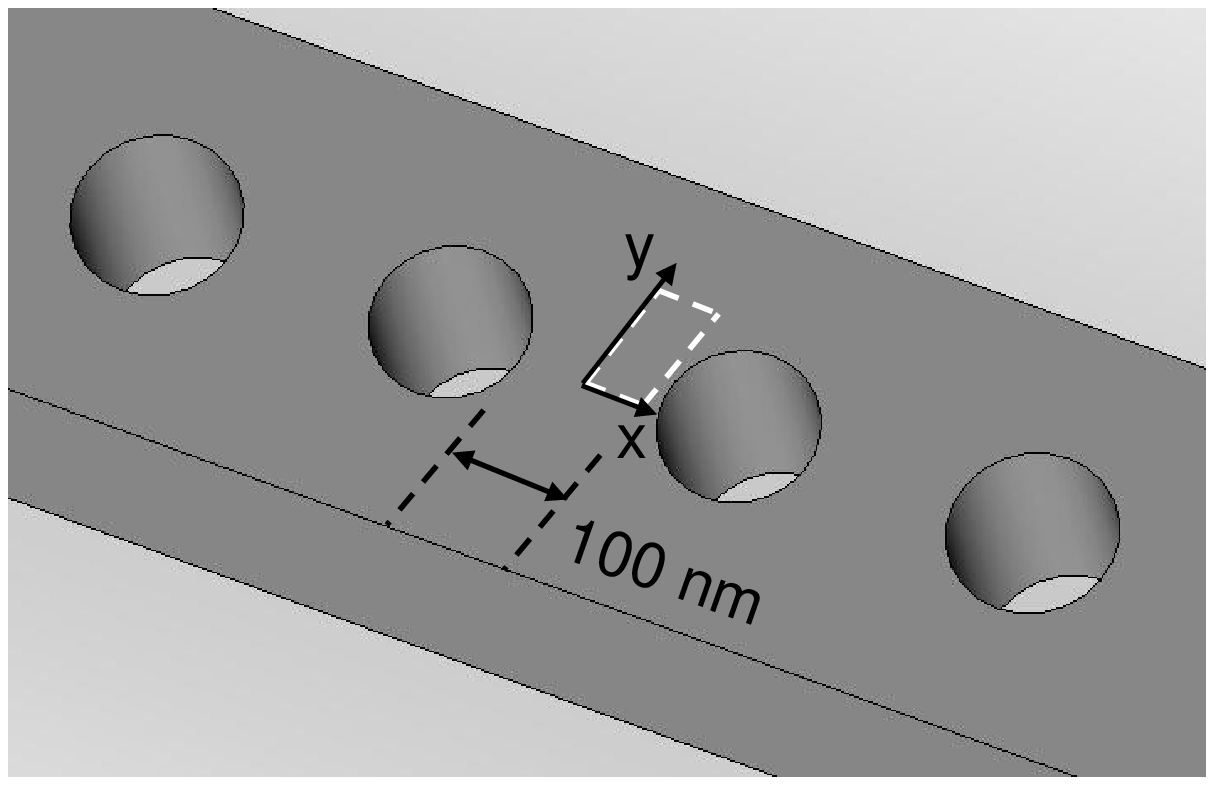} &
\includegraphics[width=6cm]{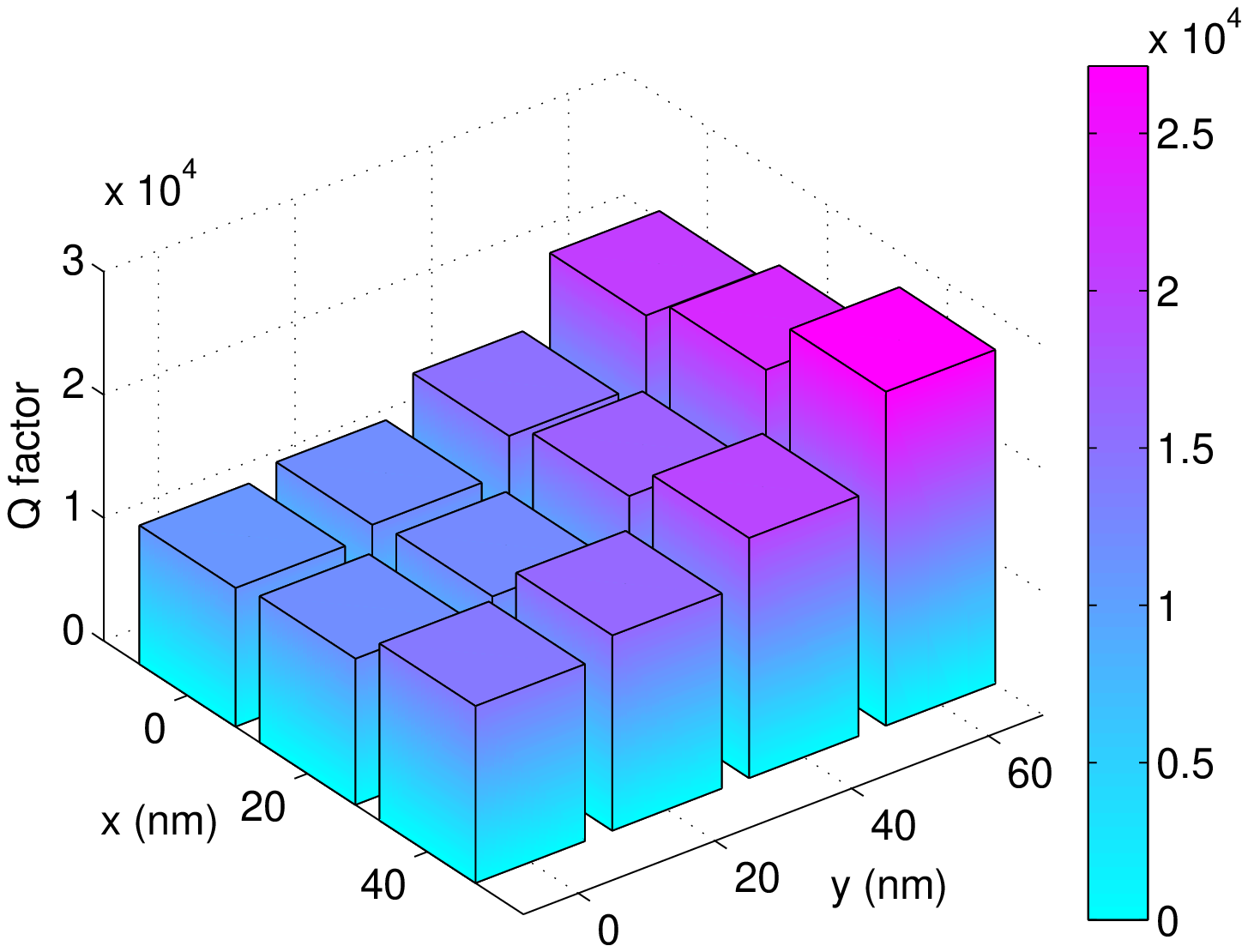} \\
(a) & (b) \\
\includegraphics[width=6cm]{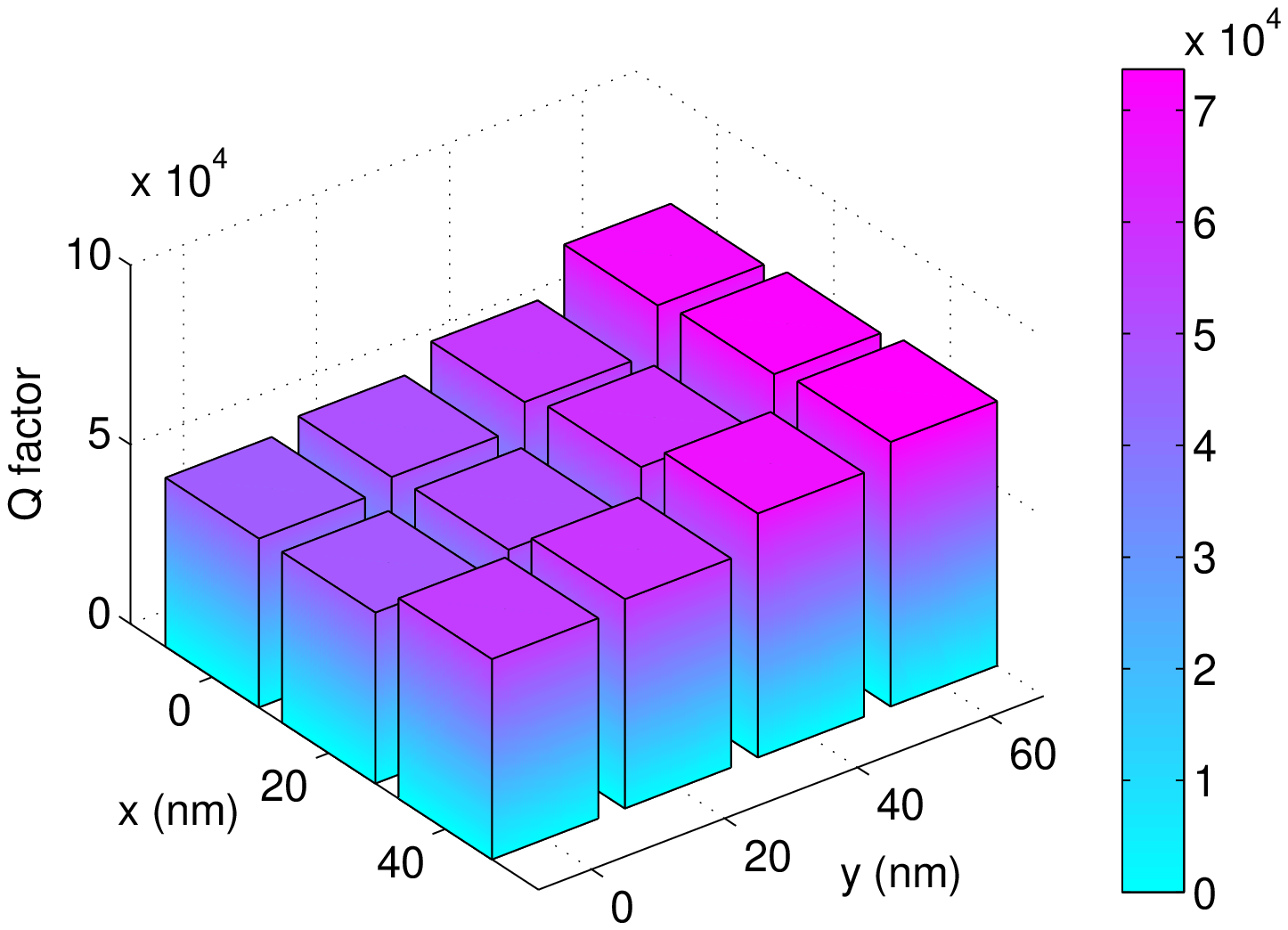} &
\includegraphics[width=6cm]{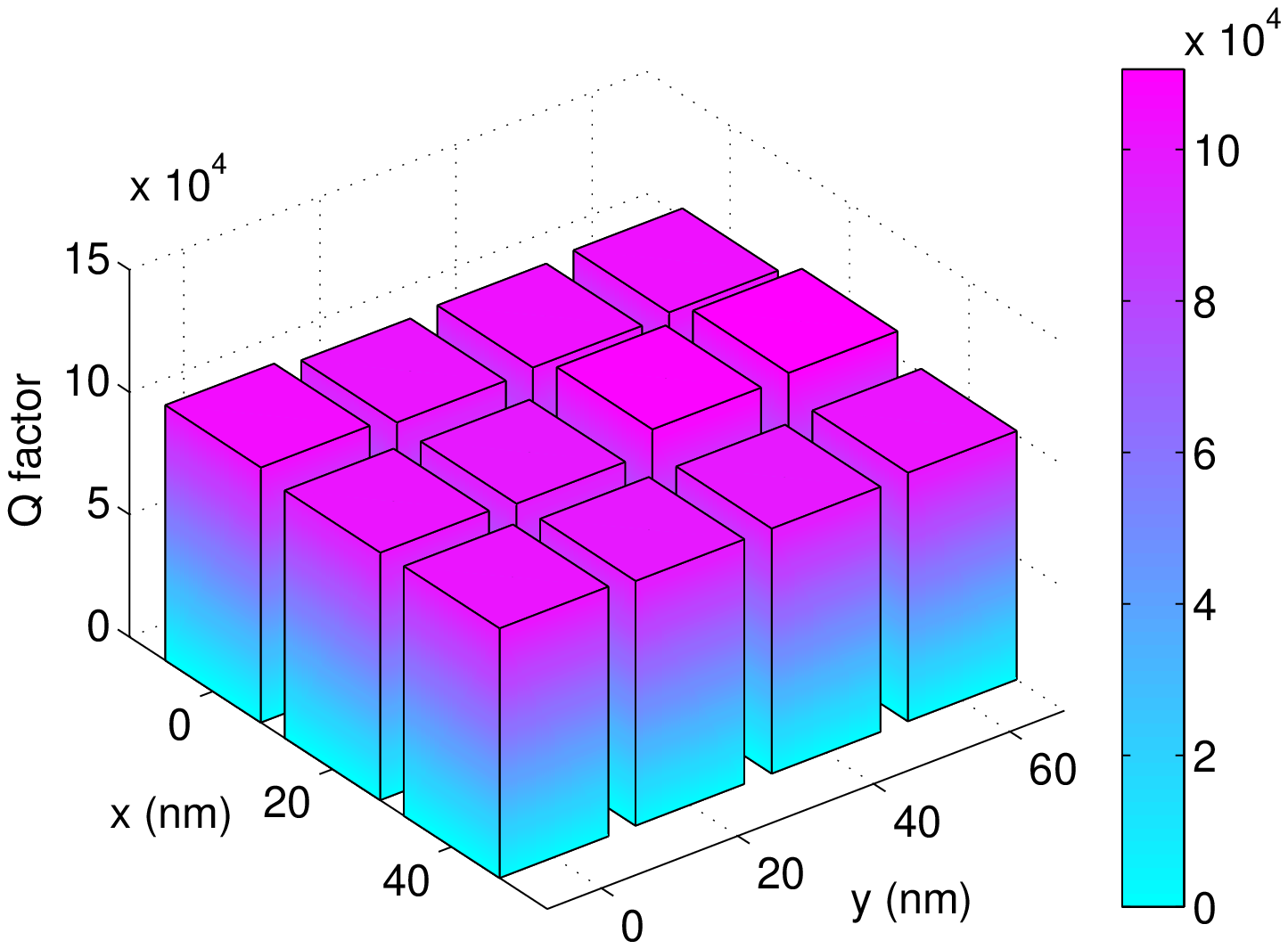} \\
(c) & (d) \\
\end{array}$
\end{center}
\caption{\label{fig:nvpos}Effect on the mode $Q$ factor
of a single diamond nanocrystal cube ($n$ = 2.43) of varying size
and displacement placed on top of the $s = 100$ nm nanocavity, which has
a maximum unperturbed $Q$ of this cavity is 115,000 (see 
Fig.~\ref{fig:QV}). The white box in (a) shows the range over which the 
nanocrystal is positioned, and the bar images show the $Q$ factor as
a function of position for NCs of size (b) 60 nm (c) 40 nm and (d) 20 nm.
($x,y$) = (0,0) denotes the nanocavity center (top surface).}
\end{figure}

The simulations are repeated for a range of NC positions to
model the effect of imperfect NC placement, and the results are 
summarized in Figure~\ref{fig:nvpos}.  The NC position is varied
over a displacement of 40 nm in the x-direction and 60 nm in the 
y-direction with respect to the center of the cavity, as shown
by the white box in Fig.~\ref{fig:nvpos}(a).  

The results show that while the relatively large 60 nm diamond NCs
have a significant impact on the mode $Q$, the situation is very
promising for 40 nm and 20 nm sized NCs.  The on-center $Q$ factor
is 47,100 for a 40 nm NC, and as the position is moved away in either 
the $x$ or $y$ directions from the mode center, the $Q$ factor increases
to a maximum of almost 74,000 at ($x,y$)= (40,60) nm.  Of course,
for positions away from the central anti-node of the mode, the
electric field strength is lower, reducing the coupling of the NC
to the cavity.  There will thus be a trade-off between $Q$ and field 
strength.

For a 20 nm
diamond NC, the $Q$ factor is close to $10^5$, or about 90\% of the
maximum unperturbed $Q$ of the cavity, regardless of NC position.
We note that Jelezko et al.~\cite{Jelezko04} observed 1.5 to 2
$\mu$s spin coherence times in 20 nm sized nanocrystallites of 
diamond.  Our results show that an NC of this size has little
effect on our cavity mode $Q$ factor whether it is embedded in
the middle or positioned on top, and therefore is a very promising 
candidate for cavity QED experiments.  

\section{Cavity QED analysis}

To achieve strong coupling between a diamond NV center and a
cavity, the coherent interaction rate, or vacuum Rabi frequency
$g$, must exceed the decoherence rates due to the
cavity loss, $\kappa$, and the spontaneous emission rate, $\gamma$;
i.e., $g > \kappa, \gamma$.  The Rabi frequency is defined as
$g = \vec{\mu} \cdot \vec{E}_{\rm 1 ph}/\hbar$, where $E_{\rm 1 ph}$
is the single photon electric field strength.  At the electric field
maximum of the cavity mode, 
\begin{equation}
E_{\rm 1 ph} = \sqrt{\frac{\hbar \omega}{2\epsilon_0 n^2 V}},
\label{eq:E1ph}
\end{equation}
where $\omega$ is the mode frequency, $V$ is the mode volume, and
$n=n_c$ is the cavity refractive index.

The dipole moment of the NV center can be deduced from its spontaneous
emission lifetime, $\tau \sim 20$ ns~\cite{Beveratos}, by using 
Fermi's golden rule for an electric dipole 
transition~\cite{Gerard,Barclay_thesis}:
\begin{equation}
\frac{1}{\tau} = \frac{2\pi}{\hbar^2} \rho(\omega) 
\langle |\langle \vec{\mu}\cdot \vec{E}_e\rangle|^2\rangle.
\end{equation}
The squared matrix element is averaged over the available emission modes.
Here, $\vec{E}_e$ is the single photon electron field at the emitter, and
$\rho(\omega)$ is the density of states in a homogeneous medium
with refractive index $n_e$: 
\begin{equation}
\rho(\omega)=\frac{\omega^2 n_e^3 V}{\pi^2 c^3}.
\end{equation}
In our context, $n_e = 2.43$ is the refractive index of the diamond nanocrystal.

Solving for $\mu$ with the help of eq.~(\ref{eq:E1ph}) for $E_e$
(with $n = n_e$) gives
\begin{equation}
\mu= \sqrt{\frac{3\pi\epsilon_0 c^3 \hbar}{\tau n_e \omega^3}}.
\label{eq:mu}
\end{equation}
The extra factor of 3 arises from the 1/3 averaging of the
squared dipolar matrix element over the random polarization of free-space
modes.  Note that this dipole moment is integrated over the entire spectrum 
of the NV emission, a point we shall return to below.

The Rabi frequency can now be evaluated from equations~\ref{eq:mu}
and~\ref{eq:E1ph} to give
\begin{equation}
g_0 = \sqrt{\frac{3\pi c^3}{2\tau\omega^2 n_e n_c^2 V}}.
\label{eq:g}
\end{equation}
The label $g_0$ indicates that this is the maximum coherent interaction
rate, which holds if the NV center is positioned at the mode maximum and 
aligned with its dipole moment parallel to the field vector.
 
When the emitter is not located at the mode maximum - e.g. if it is
positioned on top of the cavity, as sketched in Fig.~\ref{fig:3dview} 
-  $g_0$ must be scaled by $\eta = E_{\rm NV}/E_m$, 
the relative strength of the electric field 
at the NV location ($E_{\rm NV}$) compared to at the mode maximum ($E_m$).
This factor 
can be obtained from our FDTD simulations, and for a 20 nm NC on top of the 
$s=95$ nm cavity, it is given by $\eta = 0.24$.


If the NV center is embedded at the center of the cavity where the
mode peaks, there is still a reduced field strength inside the NC 
due to the larger dielectric 
constant of the diamond compared to the surrounding cavity material.
Maxwell's equations dictate that
the normal component of the electric field at the interface between
two materials (labeled 1 and 2) satisfies the boundary condition
$\epsilon_1 E_1 = \epsilon_2 E_2$.  
For the small perturbation posed by a sub-wavelength sized diamond 
nanocrystal, this picture will be more complicated than for a single
interface between two bulk dielectrics (in which case the field
would be scaled down by a factor ($2/2.43$)$^2 = 0.68$ in the higher
index diamond NC).  We can again determine $\eta$ precisely from 
FDTD simulations, and for a 20 nm diamond NC at the middle of the cavity,
$\eta = 0.85$.  


For each cavity (see e.g. Fig.~\ref{fig:QV},
we are now in a position to calculate the relevant cavity QED figures of merit, 
given by the three angular frequencies:
\begin{eqnarray}
g & = & \eta g_0, \\
\kappa & = & \frac{\omega}{2Q}, \\
\gamma & = & 2\pi/20 {\rm ns}.
\end{eqnarray}

In a diamond NV center, the zero-phonon line only contributes about 5\% of
the total emission, the rest being emitted into the phonon 
sideband~\cite{Kurtsiefer}.  Therefore,
only $\approx$ 5\% of the total emission is coupled to the cavity mode,
and $g$ must be scaled by $1/\sqrt{20}$.

\begin{figure}[tb]
\centering
\includegraphics[width=10cm]{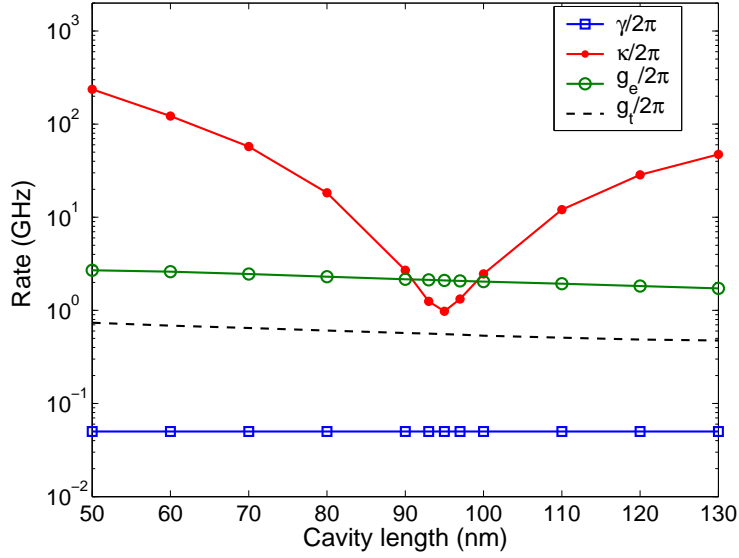}
\caption{Relevant cavity QED rates as a function of cavity length.  
$\gamma/2\pi$
is the spontaneous emission rate of the NV center, $\kappa/2\pi$ is the 
cavity decoherence rate, and $g_e/2\pi$ ($g_t/2\pi$) 
is the Rabi frequency of the 
diamond NC embedded in the center (placed on top) of the cavity.}
\label{fig:strcouple}
\end{figure}

These parameters are plotted in Figure~\ref{fig:strcouple}.  The spontaneous
emission rate $\gamma$ is independent of cavity length.  The Rabi frequency
has a weak dependence on cavity length through $\omega_0$ and  
$V$, both of which change slowly with length.  
The cavity field decay rate, $\kappa$, obviously depends strongly 
on cavity length, as it is proportional to $1/Q$.  When 
the NC is placed on top of the cavity (black dashed line), 
the condition $g > \kappa,\gamma$
is not quite satisfied for the highest $Q$ cavity.  When the NC is embedded
in the cavity (green line), however, 
the strong coupling condition is satisfied for
cavities with lengths $s = 90-100$ nm.  At 95 nm, the single-atom cooperativity
for an embedded NC is $C_e = g_e^2/\kappa\gamma = 90$, which is a promising 
figure-of-merit describing the strength of the matter-field interaction. For
the NC on top of the cavity, $C_t = 6$.

In the weak coupling, or ``bad cavity'' regime, in which the rate of 
cavity decoherence $\kappa$ exceeds the coupling rate $g_0$, 
the spontaneous emission rate will
be strongly enhanced by the Purcell effect.  At a temperature
of 1.8 K, the emission linewidth of the ZPL corresponds to a $Q$ factor of a 
few $10^7$~\cite{Tamarat}, which is much narrower than the cavity line, 
signifying that the full ZPL can be strongly enhanced 
given optimal coupling with the cavity mode.  
For the 95 nm cavity with an embedded diamond NC, the Purcell factor 
$F_p = 3 Q \eta^2 (\lambda/n)^3/4\pi^2 V = 27000$, assuming the full 
spectral and polarization alignment of the NV transition dipole
moment with the anti-node of the cavity mode.  The relative field 
strength $\eta$ at the NC position enters as a squared 
factor~\cite{Gerard}. If the NC is positioned on top of the
95 nm cavity, the maximum attainable Purcell factor is about 1,800. 
The cavity is thus a 
highly promising device to enhance the photon production rate from an NV 
center.


\section{Conclusions}

In this paper, we have engineered a high-$Q/V$ SiN$_x$ photonic crystal 
nanocavity with $n=2.0$ for the purpose of strongly coupling the 
cavity mode 
with a single NV center in a 20 nm diamond nanocrystal.
The structure should be relatively straightforward to fabricate, as
the process is based on well-known
nanofabrication techniques, and it
naturally integrates the cavity with an on-chip ridge waveguide,
allowing a well-defined output channel to be engineered for photons 
confined in the cavity mode~\cite{Banaee07}.  
This could be achieved either by 
shortening the photonic mirror to decrease the in-waveguide $Q$ 
factor, or by applying a $Q$-spoiling pulse to inject free carriers
into the cavity, as has been demonstrated in recent work on dynamic 
perturbations in photonic nanocavities~\cite{Tanabe07, McCutcheon_OE}.
We have demonstrated a cavity $Q$ factor of 230,000 with an effective 
mode volume of 0.55 ($\lambda/n$)$^3$, 
and shown that a 20 nm diamond nanocrystal located
on the cavity surface reduces the $Q$ by only $\sim 10\%$.  The same
NC embedded in the middle of the cavity increases the $Q$ factor
while reducing the mode volume.  By calculating
the Rabi frequency 
and comparing this to the decoherence rates of the system, we have
shown that the cavity with an embedded NC 
can operate in the strong coupling regime for 
realistic parameters.  We anticipate these results will open new
avenues for photonic crystal-based visible photonics in both classical
and quantum domains.

\section*{Acknowledgements}

The authors would like to thank Darrick E. Chang for his insightful comments.
Murray McCutcheon would like to thank the Natural Science and Engineering 
Research Council of Canada for their generous support.  
This work is supported in part by Harvard's National Science and Engineering 
Center (http://www.nsec.harvard.edu) and NSF NIRT grant ECCS- 0708905.

\begin{thebibliography}{99}
\bibliographystyle{osajnl}
\newcommand{\enquote}[1]{``#1''}

\bibitem{Miller}
R.~Miller, T.~E. Northrup, K.~M. Birnbaum, A.~Boca, A.~D. Boozer, and H.~J.
  Kimble, \enquote{Trapped atoms in cavity {QED}: coupling quantized light and
  matter,} Journal of Physics B \textbf{38}, S551--S565 (2005).

\bibitem{Walther}
H.~Walther, B.~T.~H. Varcoe, B.-G. Englert, and T.~Becker, \enquote{Cavity
  quantum electrodynamics,} Reports on Progress in Physics \textbf{69},
  1325--1382 (2006).

\bibitem{Weisbuch92}
C.~Weisbuch, M.~Nishioka, A.~Ishikawa, and Y.~Arakawa, \enquote{Observation of
  the coupled exciton-photon mode splitting in a semiconductor quantum
  microcavity,} Phys.\ Rev.\ Lett. \textbf{69}, 3314--3317 (1992).

\bibitem{Yoshie04}
T.~Yoshie, A.~Scherer, J.~Hendrickson, G.~Khitrova, H.~M. Gibbs, G.~Rupper,
  C.~Ell, O.~B. Shchekin, and D.~G. Deppe, \enquote{Vacuum {R}abi splitting
  with a single quantum dot in a photonic crystal nanocavity,} Nature
  \textbf{432}, 200--203 (2004).

\bibitem{Reithmaier}
J.~P. Reithmaier, G.~Sek, A.~Loffler, C.~Hofmann, S.~Kuhn, S.~Reitzenstein,
  L.~V. Keldysh, V.~D. Kulakovskii, T.~L. Reinecke, and A.~Forchel,
  \enquote{Strong coupling in a single quantum dot-semiconductor microcavity
  system,} Nature \textbf{432}, 197--200 (2004).

\bibitem{Hennessy07}
K.~Hennessy, A.~Badolato, M.~Winger, D.~Gerace, M.~Atature, S.~Gulde, S.~Falt,
  E.~L. Hu, and A.~Imamoglu, \enquote{Quantum nature of a strongly coupled
  single quantum dot-cavity system,} Nature \textbf{445}, 896--899 (2007).

\bibitem{Srinivasan07}
K.~Srinivasan and O.~Painter, \enquote{Linear and nonlinear optical
  spectroscopy of a strongly coupled microdisk-quantum dot system,} Nature
  \textbf{450}, 862--865 (2007).

\bibitem{Englund07}
D.~Englund, A.~Faraon, I.~Fushman, N.~Stoltz, P.~Petroff, and
  J.~Vu\v{c}kovi\'{c}, \enquote{Controlling cavity reflectivity with a single
  quantum dot,} Nature \textbf{450}, 857--861 (2007).

\bibitem{Childress_Science06}
L.~Childress, M.~V.~G. Dutt, J.~M. Taylor, A.~S. Zibrov, F.~Jelezko,
  J.~Wrachtrup, P.~R. Hemmer, and M.~D. Lukin, \enquote{Coherent dynamics of
  coupled electron and nuclear spins in diamond,} Science \textbf{314},
  281--285 (2006).

\bibitem{Kurtsiefer}
C.~Kurtsiefer, S.~Mayer, P.~Zarda, and H.~Weinfurter, \enquote{Stable
  solid-state source of single photons,} Phys.\ Rev.\ Lett. \textbf{85},
  290--293 (2000).

\bibitem{Gaebel06}
T.~Gaebel, M.~Domhan, I.~Popa, C.~Wittmann, P.~Neumann, F.~Jelezko, J.~R.
  Rabeau, N.~Stavrias, A.~D. Greentree, S.~Prawer, J.~Meijer, J.~Twamley, P.~R.
  Hemmer, and J.~Wachtrup, \enquote{Room-temperature coherent coupling of
  single spins in diamond,} Nature Physics \textbf{2}, 408--413 (2006).

\bibitem{Dutt}
M.~V.~G. Dutt, L.~Childress, L.~Jiang, E.~Togan, J.~Maze, F.~Jelezko, A.~S.
  Zibrov, P.~R. Hemmer, and M.~D. Lukin, \enquote{Quantum register based on
  individual electronic and nuclear spin qubits in diamond,} Science
  \textbf{316}, 1312--1316 (2007).

\bibitem{Santori}
C.~Santori, P.~Tamarat, P.~Neumann, J.~Wrachtrup, D.~Fattal, R.~G. Beausoleil,
  J.~Rabeau, P.~Olivero, A.~D. Greentree, S.~Prawer, F.~Jelezko, and P.~Hemmer,
  \enquote{Coherent population trapping of single spins in diamond under
  optical excitation,} Phys.\ Rev.\ Lett. \textbf{97}, 247401 (2006).

\bibitem{Hanson}
R.~Hanson, F.~M. Mendoza, R.~J. Epstein, and D.~D. Awschalom,
  \enquote{Polarization and readout of coupled single spins in diamond,} Phys.\
  Rev.\ Lett. \textbf{97}, 087601 (2006).

\bibitem{Meijer}
J.~Meijer, B.~Burchard, M.~Domhan, C.~Wittmann, T.~Gaebel, I.~Popa, F.~Jelezko,
  and J.~Wrachtrup, \enquote{Generation of single color centers by focused
  nitrogen implantation,} Appl.\ Phys.\ Lett. \textbf{87}, 261909 (2005).

\bibitem{Jelezko04}
F.~Jelezko, T.~Gaebel, I.~Popa, M.~Domhan, A.~Gruber, and J.~Wrachtrup,
  \enquote{Observation of coherent oscillations in a single electron spin,}
  Phys.\ Rev.\ Lett. \textbf{92}, 076401 (2004).

\bibitem{Greentree_PRA06}
A.~D. Greentree, J.~Salzman, S.~Prawer, and L.~C.~L. Hollenberg,
  \enquote{Quantum gate for ${Q}$ switching in monolithic photonic-band-gap
  cavities containing two-level atoms,} Phys.\ Rev.\ A \textbf{73}, 013818
  (2006).

\bibitem{Cirac}
J.~I. Cirac, P.~Zoller, H.~J. Kimble, and H.~Mabuchi, \enquote{Quantum state
  transfer and entanglement distribution among distant nodes in a quantum
  network,} Phys.\ Rev.\ Lett. \textbf{78}, 3221--3224 (1997).

\bibitem{vanEnk}
S.~J. van Enk, J.~I. Cirac, and P.~Zoller, \enquote{Ideal quantum communication
  over noisy channels: A quantum optical implementation,} Phys.\ Rev.\ Lett.
  \textbf{78}, 4293--4296 (1997).

\bibitem{Park06}
Y.-S. Park, A.~K. Cook, and H.~Wang, \enquote{Cavity {QED} with diamond
  nanocrystals and silica microspheres,} Nano Letters \textbf{6}, 2075--2079
  (2006).

\bibitem{Tomljenovic}
S.~Tomljenovic-Hanic, M.~J. Steel, and C.~M. de~Sterke, \enquote{Diamond based
  photonic crystal microcavities,} Opt. Express \textbf{14}, 3556--3562 (2006).

\bibitem{Kreuzer}
C.~Kreuzer, J.~Riedrich-Moller, E.~Neu, and C.~Becher, \enquote{Design of
  photonic crystal microcavities in diamond films,} Opt. Express \textbf{16},
  1632--1644 (2008).

\bibitem{Bayn}
I.~Bayn and J.~Salzman, \enquote{Utra high-{$Q$} photonic crystal nanocavity
  design: The effect of a low-index slab material,} Opt. Express \textbf{16},
  4972 (2008).

\bibitem{WangAPL07}
C.~F. Wang, R.~Hanson, D.~D. Awschalom, E.~L. Hu, T.~Feygelson, J.~Yang, and
  J.~E. Butler, \enquote{Fabrication and characterization of two-dimensional
  photonic crystal microcavities in nanocrystalline diamond,} Appl.\ Phys.\
  Lett. \textbf{91}, 201112 (2007).

\bibitem{Rivoire}
K.~Rivoire, A.~Faraon, and J.~Vu\v{c}kovi\'{c}, \enquote{Gallium phosphide
  photonic crystal nanocavities in the visible,} Appl.\ Phys.\ Lett.
  \textbf{93}, 063103 (2008).

\bibitem{Barclay06}
P.~E. Barclay, K.~Srinivasan, O.~Painter, B.~Lev, and H.~Mabuchi,
  \enquote{Integration of fiber-coupled high-{$Q$} {SiN$_x$} microdisks with
  atom chips,} Appl.\ Phys.\ Lett. \textbf{13}, 801 (2005).

\bibitem{Eichenfeld}
M.~Eichenfeld, C.~P. Michael, R.~Perahia, and O.~Painter, \enquote{Actuation of
  micro-optomechanical systems via cavity-enhanced optical dipole forces,}
  Nature Photonics \textbf{1}, 416--422 (2007).

\bibitem{Noda05}
B.~S. Song, S.~Noda, T.~Asano, and Y.~Akahane, \enquote{Ultra-high-{$Q$}
  photonic double-heterostructure nanocavity,} Nature Materials \textbf{4},
  207--210 (2005).

\bibitem{Kuramochi06}
E.~Kuramochi, M.~Notomi, S.~Mitsugi, A.~Shinya, T.~Tanabe, and T.~Watanabe,
  \enquote{Ultrahigh-{$Q$} photonic crystal nanocavities realized by the local
  width modulation of a line defect,} Appl.\ Phys.\ Lett. \textbf{88}, 041112
  (2006).

\bibitem{Barth08}
M.~Barth, N.~Nusse, J.~Stingl, B.~Lochel, and O.~Benson, \enquote{Emission
  properties of high-{$Q$} silicon nitride photonic crystal heterostructure
  cavities,} Opt. Express \textbf{93}, 021112 (2008).

\bibitem{Lalanne_JQE}
P.~Lalanne and J.~P. Hugonin, \enquote{Bloch-wave engineering for high-q,
  small-v microcavities,} IEEE J.\ Quant.\ Elec. \textbf{39}, 1430--1438
  (2003).

\bibitem{Lalanne_04}
P.~Lalanne and S.~M. J.~P. Hugonin, \enquote{Two physical mechanisms for
  boosting the quality factor to cavity volume ratio of photonic crystal
  microcavities,} Opt. Express \textbf{12}, 458--467 (2004).

\bibitem{Sauvan}
C.~Sauvan, G.~Lecamp, P.~Lalanne, and J.~Hugonin, \enquote{Modal-reflectivity
  enhancement by geometry tuning in photonic crystal microcavities,} Opt.
  Express \textbf{13}, 245--255 (2005).

\bibitem{Barth07}
M.~Barth, J.~Kouba, J.~Stingl, B.~Lochel, and O.~Benson, \enquote{Modification
  of visible spontaneous emission with silicon nitride photonic crystal
  nanocavities,} Opt. Express \textbf{15}, 17231--17240 (2007).

\bibitem{Notomi_1D}
M.~Notomi, E.~Kuramochi, and H.~Taniyama, \enquote{Ultrahigh-${Q}$ nanocavity
  with 1d photonic gap,} Opt. Express \textbf{16}, 11095--11102 (2008).

\bibitem{Lumerical}
W.~use Lumerical FDTD Solutions for all~our simulations .

\bibitem{Borselli05}
M.~Borselli, T.~J. Johnson, and O.~Painter, \enquote{Beyond the rayleigh
  scattering limit in high-{$Q$} silicon microdisks: theory and experiment,}
  Opt. Express \textbf{13}, 1515--1530 (2005).

\bibitem{Zhang}
Y.~Zhang and M.~Lon\v{c}ar, \enquote{Ultra-high quality factor optical
  nanocavities based on semiconductor nanowires,} submitted to Opt. Express
  (2008).

\bibitem{Srinivasan_OE}
K.~Srinivasan and O.~Painter, \enquote{Momentum space design of high-${Q}$
  photonic crystal optical cavities,} Opt. Express \textbf{15}, 670--684
  (2002).

\bibitem{Painter99}
O.~Painter, J.~Vu\v{c}kovi\'{c}, and A.~Scherer, \enquote{Defect modes of a
  two-dimensional photonic crystal in an optically thin dielectric slab,} J.\
  Opt.\ Soc.\ Am.\ B \textbf{16}, 275--285 (1999).

\bibitem{Zain}
A.~R.~M. Zain, M.~Gnan, H.~M.~H. Chong, M.~Sorel, and R.~M. D.~L. Rue,
  \enquote{Tapered photonic crystal microcavities embedded in photonic wire
  waveguides with large resonance quality-factor and high transmission,} IEEE
  Phot. Tech. Lett. \textbf{20}, 6--8 (2008).

\bibitem{Hennessy04}
K.~Hennessy, A.~Badolato, P.~M. Petroff, and E.~L. Hu, \enquote{Positioning
  photonic crystal cavities to single {I}n{A}s quantum dots,} Photonics and
  Nanostructures \textbf{2}, 65--72 (2004).

\bibitem{Badolato05}
A.~Badolato, K.~Hennessy, M.~Atature, J.~Dreiser, E.~Hu, P.~M. Petroff, and
  A.~Imamoglu, \enquote{Deterministic coupling of single quantum dots to single
  nanocavity modes,} Science \textbf{308}, 1158--1161 (2005).

\bibitem{Koenderink}
A.~F. Koenderink, M.~Kafesaki, C.~M. Soukoulis, and V.~Sandoghdar,
  \enquote{Spontaneous emission in the near field of two-dimensional photonic
  crystals,} Optics Letters \textbf{30}, 3210--3212 (2005).

\bibitem{Beveratos}
A.~Bevaratos, S.~Kuhn, R.~Brouri, T.~Gacoin, J.-P. Poizat, and P.~Grangier,
  \enquote{Room temperature stable single-photon source,} Eur.\ Phys.\ J.\ D
  \textbf{18}, 191--196 (2002).

\bibitem{Gerard}
J.-M. G\'{e}rard, \emph{Single quantum dots: fundamentals, applications, and
  new concepts, P. Michler (ed.)} (Springer, 2003), chap. Solid-state
  cavity-quantum electrodynamics with self-assembled quantum dots, pp.
  269--314.

\bibitem{Barclay_thesis}
P.~E. Barclay, \enquote{Fiber-coupled nanophotonic devices for nonlinear optics
  and cavity {QED},} PhD. Thesis, California Institute of Technology  (2007).

\bibitem{Tamarat}
P.~Tamarat, T.~Gaebel, J.~R. Rabeau, M.~Khan, A.~D. Greentree, H.~Wilson,
  L.~C.~L. Hollenberg, S.~Prawer, P.~Hemmer, F.~Jelezko, and J.~Wrachtrup,
  \enquote{Stark shift control of single optical centers in diamond,} Phys.\
  Rev.\ Lett. \textbf{97}, 083002 (2006).

\bibitem{Banaee07}
M.~G. Banaee, A.~G. Pattantyus-Abraham, M.~W. McCutcheon, G.~W. Rieger, and
  J.~F. Young, \enquote{Efficient coupling of photonic crystal microcavity
  modes to a ridge waveguide,} Appl.\ Phys.\ Lett. \textbf{90}, 193106 (2007).

\bibitem{Tanabe07}
T.~Tanabe, M.~Notomi, E.~Kuramochi, A.~Shinya, and H.~Taniyama,
  \enquote{Trapping and delaying photons for one nanosecond in an ultrasmall
  high-{$Q$} photonic-crystal nanocavity,} Nature Photonics \textbf{1}, 49--52
  (2007).

\bibitem{McCutcheon_OE}
M.~W. McCutcheon, A.~G. Pattantyus-Abraham, G.~W. Rieger, and J.~F. Young,
  \enquote{Emission spectrum of electromagnetic energy stored in a dynamically
  perturbed optical microcavity,} Opt. Express \textbf{15}, 11472--11480
  (2007).


\end{thebibliography}
\end{document}